\documentclass[a4paper,11pt]{article}
\pdfoutput=1 % if your are submitting a pdflatex (i.e. if you have
\usepackage{jcappub} % for details on the use of the package, please
\usepackage[T1]{fontenc} % if needed

\usepackage{multicol}
\usepackage{subcaption}

\title{\boldmath A comparative study of cosmological constraints from weak lensing using Convolutional Neural Networks}

\author[a,b]{Divij Sharma}
\author[a,b]{Biwei Dai}
\author[a,b]{Uroš Seljak}

\affiliation[a]{Berkeley Center for Cosmological Physics and Department of Physics, University of California, Berkeley, CA 94720, USA}
\affiliation[b]{Lawrence Berkeley National Lab, 1 Cyclotron Road, Berkeley, CA 94720, USA}

\emailAdd{divijsharma@berkeley.edu}
\emailAdd{biwei@berkeley.edu}
\emailAdd{useljak@berkeley.edu}

\abstract{Weak Lensing (WL) surveys are reaching unprecedented depths, enabling the investigation of very small angular scales. At these scales, nonlinear gravitational effects lead to higher-order correlations making the matter distribution highly non-Gaussian. Extracting this information using traditional statistics has proven difficult, and Machine Learning based summary statistics have emerged as a powerful alternative. We explore the capabilities of a discriminative, Convolutional Neural Networks (CNN) based approach, focusing on parameter constraints in the ($\Omega_m$, $\sigma_8$) cosmological parameter space. Leveraging novel training loss functions and network representations on WL mock datasets without baryons, we show that our models achieve $\sim 5$ times stronger constraints than the power spectrum, $\sim 3$ stronger constraints than peak counts, and $\sim 2$ stronger constraints than previous CNN-learned summary statistics and scattering transforms, for noise levels relevant to Rubin or Euclid. For WL convergence maps with baryonic physics, our models achieve $\sim 2.3$ times stronger constraining power than the power spectrum at these noise levels, also outperforming previous summary statistics. To further explore the possibilities of CNNs for this task, we also discuss transfer learning where we adapt pre-trained models, trained on different tasks or datasets, for cosmological inference, finding that these do not improve the performance.}

\begin{document}
\maketitle
\flushbottom

%%%%%%%%%%%%%%%%%%%%%%%%%%%%%%%%%%%%%%%%%%%%%%%%%%

%%%%%%%%%%%%%%%%% BODY OF PAPER %%%%%%%%%%%%%%%%%%

\section{Introduction}\label{sec:Introduction}
Weak gravitational lensing (WL) is the distortion of light from distant galaxies caused by the gravitational influence of intervening large-scale structures that trace total matter in the universe. This phenomenon creates a subtle cosmic shear pattern in the sky, altering the observed shapes and orientations of background galaxies. The distortion of galaxy shapes, quantified through summary statistics, holds valuable information about underlying cosmological parameters \citep{Bartelmann_2001, Kilbinger_2015}. Various surveys, such as the \href{https://www.darkenergysurvey.org/}{Dark Energy Survey} (DES), \href{https://hsc.mtk.nao.ac.jp/ssp/}{Hyper Suprime-Cam Survey} (HSC), \href{https://sci.esa.int/web/Euclid}{Euclid}, the \href{https://lsst.org}{Vera Rubin Observatory} (Rubin), and the \href{https://roman.gsfc.nasa.gov/}{Nancy Grace Roman Space Telescope} (Roman), aim to map this cosmic shear across large areas of the sky. These surveys will provide observational data that can be used to constrain fundamental cosmological parameters, particularly $\Omega_m$ (matter density) and $\sigma_8$ (amplitude of matter fluctuations) that WL signals are most sensitive to in the standard cosmological model \citep{Joudaki_2016, Kohlinger_2017, Hikage_2019, Hamana_2020}.

To extract information from WL data, various summary statistics are employed. At the two-point level, WL data are analyzed using correlation functions such as the power spectrum of the shear or convergence. However, these traditional summary statistics leave a wealth of information untapped in the WL signal due to the highly non-Gaussian features at small scales. To address this limitation, various other summary statistics have emerged, such as higher-order correlation functions \citep{Takada_2002, Takada_2003, Schneider_2005, 
Zaldarriaga_2003, peebles1980, peebles_groth, Sefusatti_2006, Semboloni_2010, Fu_2014}. These are challenging because of the large number of coefficients, and the difficulty to measure the associated covariance matrix, as well as sensitivity to outliers \citep{kendall1946}. Other 
methods that have been proposed are peak counts \citep{Marian_2009, Dietrich_2010, Kratochvil_2010, Yang_2011}, Minkowski functionals \citep{Mecke_1993, Sato_2001, Guimaraes_2002, Kratochvil_2012}, scattering transform coefficients \citep{Allys_2020, Cheng_2020}, and statistics learned by neural networks \citep{Fluri_2018, Charnock_2018, Makinen_2021, Gupta_2018}.

Convolutional Neural Networks (CNNs) have been employed in various studies to estimate cosmological parameters from WL convergence maps  \citep{Gupta_2018, Ribli_2019, Fluri_2018, Fluri_2019, villaescusanavarro2020neural,Jeffrey_2020, Makinen_2021, Lu_2022, Lu_2023}. \cite{Gupta_2018} trained CNNs on noise-free convergence maps, showcasing a 5 times improvement in the precision of \( \Omega_m - \sigma_8 \) constraints compared to the power spectrum analysis. \cite{Ribli_2019} trained CNNs with a different architecture on convergence maps with different levels of shape noise, achieving a 2.4--2.8 times improvement over power spectrum in parameter constraints for surveys like Rubin.
\cite{Fluri_2019} leveraged the KiDS-450 tomographic WL dataset, illustrating a 30\% enhancement in \( S_8 = \sigma_8(\Omega_m/0.3)^{0.5} \) parameter constraints compared to power spectrum analysis. 
\cite{Matilla_2020} reported a 20\% improvement in constraints over traditional methods (power spectrum, peak counts, and Minkowski functionals) using their CNN framework.
\cite{Jeffrey_2020} 
%Additionally, \cite{villaescusanavarro2020neural} applied CNNs to Gaussian random fields and a simplified baryonic model, demonstrating CNNs' ability to extract the maximum available cosmological information in Gaussian random fields and marginalize over baryonic effects.
\cite{Lu_2022} studied the impact of baryonic effects on WL analysis with CNNs, and they further applied their framework to HSC first-year data, finding a factor of $3$ improvement in $\Omega_m$ constraints over power spectrum \citep{Lu_2023}. 

More recently, \cite{Dai2022translation} and \cite{dai2023multiscale} proposed using generative Normalizing Flows to create samples and model the field-level likelihood of weak lensing maps. On weak lensing mock datasets, \cite{dai2023multiscale} showed that Multiscale Flow (MSF) outperforms the power spectrum by factors of $3-9$ for different noise levels relevant to different surveys. It also achieves about two times higher constraining power when compared to peak counts, CNNs, and scattering transform summary statistics. 
However, CNNs are easier to train and work with than the generative Multiscale Flow model. Moreover, for cosmological inference from surveys with large survey areas, CNNs can be applied effectively by cropping the survey images and combining the summary statistics obtained from the CNNs \citep{Lu_2023}. Multiscale Flow, being a generative model with full dimensionality of the data, does not scale as well with growing survey size.

The ability of MSF to outperform current 
implementations of CNN raises the 
question whether this is inherently
due to their generative training, as 
has been suggested in previous work \citep{NgJordan}. An alternative explanation is that the existing CNN analyses have not been optimal in terms of architectural choices and training methods. 
%Residual Neural Networks (ResNets) offer promise in extracting non-Gaussian cosmological information directly from weak lensing maps due to their unique architectural design \citep{He_2016, He2016identity}. ResNets share weights across different parts of the input, enabling them to find patterns across different scales, unlike fully connected neural networks.
%ResNets have achieved state-of-the-art performance in image classification benchmarks like ImageNet, object detection, and semantic segmentation. By using residual connections, or skip connections, ResNets facilitate training very deep neural networks effectively. These connections allow the network to bypass certain layers, enabling the training of deep neural networks.  ResNets are thus a good candidate 
%to explore whether stronger constraining power can be achieved than previous CNN and other summary statistics. 
In this paper, we therefore explore the ability of CNNs to constrain cosmology using a variety of novel techniques and loss functions that have been recently proposed. We train our models on WL
convergence maps with and without baryons and compare their performance to that of the power spectrum, peak counts, scattering transform and previous CNN works. We will show that our models outperform these statistics by considerable factors.

This paper is organized as follows: Section \ref{sec:Materials} provides a comprehensive overview of the maps we used to train and test our models, the methodology employed during training, and post-training-prediction procedures to obtain parameter constraints. In Section \ref{sec:Results}, we present and discuss our results of parameter constraints derived from the trained models. Section \ref{sec:TL} explores our results employing transfer learning, where we adapt pre-trained models to predict cosmologies using the same WL maps. We summarize and conclude in section \ref{sec:Conclusions}.

\section{Materials and Methods}\label{sec:Materials}
In this section, we describe the data, models, and methods we use to get stringent parameter constraints in $\Omega_m - \sigma_8$ cosmological plane. First, we describe the weak lensing convergence maps that we use for training and testing our models. We use two different datasets for the types of convergence maps: one with only dark matter, and another dataset with baryonic physics added. Second, we describe the CNN models used and the training loss functions and methodologies we implemented during training to get varying degrees of constraining powers. Finally, we describe how we use our trained CNNs' outputs as a summary statistic for parameter inference as well as the other summary statistics that we compare our results against.

\subsection{Weak lensing maps}
\subsubsection{Dark-matter-only weak lensing maps}
The dark-matter-only (DM-only) weak lensing convergence maps utilized in this study, obtained from \cite{Gupta_2018}, stem from a suite of 80 N-body simulations characterized by spatially flat $\Lambda$CDM cosmologies. Each simulation varies in cosmological parameters $\Omega_m$ and $\sigma_8$, while maintaining fixed values for other parameters: $\Omega_b = 0.046$, $h = 0.72$, and $n_s = 0.96$. Specifically, the sampling of $\Omega_m$ and $\sigma_8$ is non-uniform, concentrating more densely around $\Omega_m = 0.26$ and $\sigma_8 = 0.8$. These simulations evolve $512^3$ dark matter particles within a $240 h^{-1}$ Mpc box, employing the N-body code Gadget-2 \citep{GADGET-2}. Snapshots are recorded between redshifts $0 < z < 1$, spaced $80 h^{-1}$ Mpc apart in comoving distance.

Weak lensing convergence maps with a field of view of $3.5 \times 3.5$ deg$^2$ are then generated by ray-tracing the snapshots of N-body simulations to redshift $z = 1$ with a multiple lens plane algorithm~\citep{Schneider_lenses}. From each simulation, $512$ pseudo-independent maps are derived by incorporating random rotations, flips, and shifts to the snapshots. Detailed insights into the map generation process can be found in ~\cite{Gupta_2018}.

In \cite{Ribli_2019}, the maps are downscaled from a resolution of $1024^2$ to $512^2$ while introducing Gaussian galaxy shape noise. We further downscale the maps from $512^2$ to $256^2$ for faster training times, without seeing considerable degradation in the constraining power of the trained models. The standard deviation $\sigma$ of the Gaussian noise is computed as 
\begin{align}
    \sigma = \frac{\sigma_\epsilon}{\sqrt{2 n_{\rm \text{g}} A_{\rm \text{pixel}}}}
\end{align}
with $\sigma_\epsilon \sim 0.4$ denoting the mean intrinsic ellipticity of galaxies, and $A_{\rm \text{pixel}}$ signifying the pixel area. This dataset incorporates three distinct galaxy density scenarios: $n_{\rm \text{g}} = 10$, $30$, and $100$ arcmin$^{-2}$. A noise level of 10 galaxies arcmin$^{-2}$ characterizes typical ground-based surveys like CFHTLens, DES, or KiDS. Around 30 galaxies arcmin$^{-2}$ represent the targeted noise level for surveys like Rubin or Euclid, while future space missions like Roman might access between 50 and 75 galaxies arcmin$^{-2}$. The scenario with 100 galaxies arcmin$^{-2}$ stands as an optimistic anticipation for forthcoming space-based surveys.
While \cite{Ribli_2019} apply a $1$ arcmin Gaussian kernel for map smoothing to augment signal-to-noise ratio (S/N) and mitigate small-scale information affected by baryonic physics, we find no need for this procedure, and our analysis does not use smoothing techniques on the noisy maps.

\subsubsection{Weak lensing maps with baryons}
To understand the influence of baryonic effects within our analysis, we analyze weak lensing convergence maps from \cite{Lu_2022}. These maps stem from an identical set of N-body simulations and ray-tracing methodologies as the previous dark-matter-only maps, with identical resolution and field of view. However, a key distinction lies in the post-processing of simulation snapshots to incorporate baryonic effects. A concise overview of this post-processing step follows herein; for more comprehensive details, readers are referred to \cite{Lu_2022, Lu_2021}.

In their work, \cite{Lu_2022} identify all dark matter halos with masses greater than a mass threshold of $10^{12} M_{\rm \odot}$ within simulation snapshots. These halos' constituent particles are then substituted with spherically symmetric analytical halo profiles, characterizing the matter distribution within the halos. The analytical halo profile, derived from the Baryon Correction Model (BCM) \citep{Aric_2020}, delineates halos through four components: the central galaxy (stars), bounded gas, ejected gas (attributed to AGN feedback), and relaxed dark matter. Parameterized by four free parameters – $M_c$ (characteristic halo mass retaining half the total gas), $M_{\rm 1,0}$ (characteristic halo mass for a galaxy mass fraction of $0.023$), $\eta$ (maximum distance of the ejected gas from the parent halo), and $\beta$ (logarithmic slope of the gas fraction vs. the halo mass). This model eliminates the substructure and non-spherical shapes of the halos, but it has been argued that the morphological disparities between simulated halos and spherical analytical profiles are statistically insignificant compared to uncertainties in the power spectrum and peak counts within an HSC-like survey~\citep{Lu_2022}.

\cite{Lu_2022} generate 2048 maps for each cosmology, each with distinct baryon parameters. They utilize the first 1024 maps for training a CNN and employ the remaining 1024 maps to calculate the mean and covariance matrix of the CNN-learned summary statistics. Aligned with their methodology, our analysis also uses the first 1024 maps for training and the subsequent 1024 maps for statistical inference.
Similar to the preprocessing steps applied to the dark-matter-only maps, we downsample these maps to a resolution of $256^2$ and introduce Gaussian shape noise. This dataset encompasses three galaxy densities: $n_g = 20$, $50$, and $100$ arcmin$^{-2}$, facilitating a more comprehensive comparison with the results of \cite{Lu_2022}.

\subsection{Neural network architectures}
In our study, we leverage the learning power of deep residual neural networks (ResNets) \citep{He_2016, He2016identity}. Specifically, we experimented with ResNet18, ResNet34, and ResNet50. These architectures have played a pivotal role in image data based deep learning over the past several years. ResNet18, ResNet34, and ResNet50 are characterized by their depth, a defining feature that allows them to capture intricate patterns and nuances in the data. The innovative element of ResNets is the introduction of residual connections, or skip connections, which facilitate the training of very deep networks. This helps overcome the vanishing gradient problem and, as a result, enables the training of neural networks with many layers. ResNet18, the smallest of the three, consists of 18 convolution layers,  ResNet34 is deeper with 34 layers, with ResNet50 being even deeper, featuring 50 layers. These networks have demonstrated remarkable performance across a wide range of computer vision tasks, making them invaluable tools for image based data such as weak lensing data.

Our empirical experiments show that the choice of neural network architecture does not significantly influence our results. We present a comparison between different neural network architectures using the same loss function in table \ref{tab:table1}. Hence, we opted to employ ResNet18 consistently for all experiments detailed in this paper. We made this choice due to its relatively shallower depth compared to ResNet34 and ResNet50, balancing computational complexity while preserving the capacity to capture meaningful representations from our weak lensing data.

\subsection{Training}
\subsubsection{Training Loss Functions}
We utilized various loss functions to train our models, aiming to optimize the performance of our networks in capturing and interpreting the complex cosmological parameter space. We detail the loss functions in the following paragraphs. We find that CNNs achieve different constraining power depending on the choice of the loss function, as detailed in section \ref{sec:Results}.

First, we employed the Mean Squared Error (MSE) loss, defined as:

\begin{align}
    \text{MSE loss} = \frac{1}{n} \sum_{\rm i=1}^{n} (y_i - \hat{y}_i)^2
\end{align}

This loss function measures the squared differences between predicted (\(\hat{y}_i\)) and actual (\(y_i\)) values across \(n\) data points. It corresponds to maximum likelihood estimation (MLE), under the assumption that the likelihood is Gaussian with an identity covariance matrix. %While effective due to its differentiability for gradient-based optimization, the MSE loss can be sensitive to outliers, impacting its suitability for specific scenarios.

Another loss function we use involves transforming cosmological parameters into approximately uncorrelated parameters. Specifically, we introduced \(S_8\) as a function of \(\sigma_8\) and \(\Omega_m\) via \(S_8 = \sigma_8 \left(\frac{\Omega_m}{0.3}\right)^{0.5}\), coupled with an approximately orthogonal parameter \(\text{Ortho} = \frac{\sigma_8^2}{2} - \Omega_m^2\). Using an L2 loss function with custom weights, corresponding to the expected errors associated with these parameters, we formulated the following loss:
\begin{align}
\text{MSE\(_{\rm NP}\) loss} = w_1 \cdot (S_8 - S_{\rm 8,\text{true}})^2 + w_2 \cdot \left(\text{Ortho} - \text{Ortho}_{\rm \text{true}}\right)^2
\end{align}
Here, \(w_1\) and \(w_2\) denote the expected inverse variance for \(S_8\) and \(\text{Ortho}\), respectively. This loss 
has the advantage of incorporating 
non-Gaussian correlations, due to 
the nonlinear nature of the parameter
transformation. 

Next, we explored Principal Component Analysis (PCA) by transforming the original parameter space into a different space with the same dimensionality (2 for DM-only maps and 6 for baryon maps), with the goal of optimizing training parameters for the MSE loss. This transformed space, characterized by principal components capturing data variations, allows us to use an L2 loss function:
\begin{align}
\text{MSE\(_{\rm PCA}\) loss} = (\mathbf{X} - \hat{\mathbf{X}})^T \cdot \mathbf{W} \cdot (\mathbf{X} - \hat{\mathbf{X}})
\end{align}
Here, \(\mathbf{X}\) represents the parameter set in the transformed PCA space, \(\hat{\mathbf{X}}\) indicates predicted values in the same space, and \(\mathbf{W}\) symbolizes a diagonal weight matrix holding expected inverse variance values for each parameter. By utilizing PCA followed by an L2 loss, we  improve model efficiency over the regular MSE loss. However, the PCA transformation is linear and unable to account for non-Gaussian posteriors. 

Next, we integrated the Neural Posterior Estimation (NPE) methodology proposed by \cite{Jeffrey_2020, zeghal2022neural} for Simulation-Based Inference (SBI). This method aims to directly learn the posterior distribution by employing a Neural Density Estimator (NDE) like a Normalizing Flow (NF). The loss function in this context, called the  Variational Mutual Information Maximization (VMIM) loss \citep{Barber_2003} or the expected Negative Log-Likelihood (NLL) loss (\(L_{\rm \text{NLL}}\)), minimizes the expected negative log-posterior:
\begin{align}
\text{VMIM loss} = \mathbb{E}_{\rm p(\theta, x)}\left[-\log p_\phi(\theta | x)\right] = L_{\rm \text{NLL}}
\end{align}
This loss function involves compressing the WL maps using a neural network -- a ResNet in our case -- into a summary statistic, and employing conditional normalizing flows to approximate the posterior distribution $p_\phi(\theta | x)$ given learned summary statistic.

\subsection{Hyperparameters and Training Strategies}
Throughout the training phase, we conducted careful experimentation with various hyperparameters to yield optimal model performance and convergence. To prevent overfitting and regularizing the model, we used a weight decay factor of \(1 \times 10^{-3}\), after establishing this 
to be the optimal value. This factor played a crucial role in constraining the complexity of the model and preventing excessive sensitivity to noise in the training data.

Additionally, we organized the training data into batches during training, with each batch containing randomly sampled 128 WL maps. This batch size was selected to balance computational efficiency and model convergence while ensuring a manageable memory during training.

In our training process, a piecewise constant learning rate schedule was adopted to dynamically adjust the learning rate at specific milestones throughout the training iterations. This adaptive learning rate schedule was essential in controlling the rate of model parameter updates, enhancing the model's ability to navigate the loss landscape effectively and converge toward an optimal solution.

Despite the use of regularization, 
we found ResNets to overfit the 
training data. 
To further mitigate overfitting, we employed early stopping. We separated fiducial cosmology maps from the training data and designated them for validation purposes. During training iterations, the model's performance on the validation set was monitored, and the model achieving the lowest loss on this validation set was saved. This approach ensured that the model's generalizability was preserved by preventing it from becoming overly specialized to the training data and thereby improving its performance on unseen test data.

\subsection{Parameter Inference}
\label{sec:inference}
Neural network output predictions 
for cosmological parameters can be interpreted as summary statistics, akin to conventional statistics like the power spectrum or peak counts, regardless of their direct connection to the underlying parameters. This approach has been widely applied in prior studies \citep{Gupta_2018, Ribli_2019, Matilla_2020,Lu_2022,Lu_2023}. Hence, a statistic, in this context, denotes either the prediction output by a network or a combination of various statistics.
The advantage of this summary 
statistic view is that it can 
be further improved if necessary, and that its density estimation 
is relatively simple since it is 
very low dimensional. In contrast, 
high dimensional summary 
statistics suffer from the need 
to have many simulations to 
learn their covariance matrix, and 
typically one needs several times
more training data than the 
number of summary statistics. 

Following previous studies, given a summary statistic comprising \(d\) observables, we model the likelihood of observing \(y\) as a multidimensional Gaussian distribution:
\begin{equation}
\label{eq:likelihood}
p(y|\theta) \propto \frac{1}{\sqrt{\text{det }C}} \exp\left(-\frac{1}{2} [y - y(\theta)]^T C^{-1}(\theta) [y - y(\theta)]\right).
\end{equation}
Here, \(\theta\) represents the underlying cosmological and baryonic parameters, and we use \(N\) realizations to estimate the covariances \(C\). 
%, and \(A_{\rm \text{survey}}\) denotes the area of a hypothetical survey. The term \(\frac{N - d - 2}{N - 1}\) normalizes the estimation of the precision matrix \(C^{-1}\) to alleviate bias \citep{Hartlap_2007}. 
For calculating the expected values, \(y(\theta)\), and the covariances, $C(\theta)$, we follow slightly different approaches for the two different WL map datasets, which the following subsections outline.

\subsubsection{Dark Matter Only Maps}
To extend the likelihood across the entire \(\Omega_m - \sigma_8\) space from the 80 discrete cosmologies sampled by the DM-only maps, we perform linear interpolation for both the expected value, \(y(\theta)\), and the covariances. Using Bayes' theorem, we then estimate the posterior distribution of \(\theta\) considering a mock observation at the fiducial parameters \(y(\theta_0)\) while employing uniform priors on all parameters.

\subsubsection{Baryon Maps}
To extend the likelihood across the \(\Omega_m - \sigma_8\) space from the 75 discrete cosmologies sampled by the baryon maps, we undertake a two-step interpolation for the expected value \(y(\theta)\) following \cite{Lu_2022}: (1) fitting a second-degree polynomial for the four baryonic parameters within each cosmology, and (2) linear interpolation between cosmologies. We assume constant covariances across different baryonic parameters, estimating these covariances for each cosmology and interpolating them linearly. Using Bayes' theorem, we estimate the posterior distribution of \(\theta\) based on a mock observation at the fiducial parameters \(y(\theta_0)\), while adopting log-uniform priors on all parameters.

\subsection{Power Spectrum}
We take the power spectrum constraining power estimation from \cite{dai2023multiscale}, who measure the power spectrum of convergence maps using the publicly available LensTools package \citep{LensTools}. The power spectrum is calculated in 20 bins in the range \(100 \leq l \leq 37500\) with logarithmic spacing, following the settings adopted in \cite{Ribli_2019} and \cite{Cheng_2020}. The logarithm of the power spectrum is taken to be observable for parameter inference.

\subsection{Peak Count}
For peak count constraining power, we take the peak count measurement from \cite{Ribli_2019}, who identify the local maxima of convergence maps and measure the binned histogram of the peaks as a function of their \(\kappa\) value. They use 20 linearly spaced \(\kappa\) bins in total.

\subsection{Scattering Transform}
Initially introduced as a method to extract information from high-dimensional data by \cite{mallat2012group}, the scattering transform has been recently applied to cosmological data analysis. In studies \citep{Cheng_2020, Allys_2020, Cheng_2021, Valogiannis_2022}, it showcased improvements over the power spectrum, especially in low noise scenarios. This technique generates a cluster of new fields through recursive wavelet convolutions and modulus operations. The expected values of these derived fields serve as scattering coefficients and act as summary statistics. In this study, we compare our findings directly to the results obtained by \cite{Cheng_2020}, who estimated the constraining capabilities of the scattering transform using Fisher forecast techniques on the same dataset that we use.

\section{Results}\label{sec:Results}
In this section, we present the constraining power of our trained CNN models for each loss mentioned above and compare them to traditional summary statistics as well as previous CNN-learned summary statistics. First, we describe the results for the  DM-only WL maps dataset, and then we explore the effects of baryons on our constraining power using the baryon maps dataset.

\subsection{Cosmological constraints from noisy weak lensing maps}
First, we evaluate the efficacy of our trained CNNs using the $3.5 \times 3.5$ deg$^2$ mock weak lensing convergence maps from \cite{Ribli_2019} to conduct field-level inference. The performance of our model is assessed based on the constraining power, measured as the reciprocal of the 1$\sigma$ confidence area on the ($\Omega_m$, $\sigma_8$) plane. Table \ref{tab:table1} presents a comparative analysis showcasing the figure of merit of our CNNs across different noise levels, combined with results obtained using traditional summary statistics such as the power spectrum, peak count, scattering transform \citep{Cheng_2020}, and statistics derived from CNNs in \cite{Ribli_2019}.

Our models achieve the best performance among all the methods, surpassing the other methods, including the power spectrum, by substantial margins. Specifically, on galaxy densities $n_g = 10$, $30$ (relevant for Rubin or Euclid), and $100$ arcmin$^{-2}$, our CNNs outperform the power spectrum by factors of $2.4$, $5$, and $5$, respectively, using the VMIM loss. Notably, our best-performing model demonstrates a marked enhancement in constraining power, achieving $\sim 1.5-2$ times higher performance compared to peak counts, previous CNN models, and scattering transform for all noise levels. 

Using $MSE_{\rm NP}$ and $MSE_{\rm PCA}$ losses, our CNNs also outperform power spectrum analysis by factors of $\sim$ $2$, $3$, and $4$ for the three noise levels respectively. They also achieve $\sim 1.2-2$ times stronger parameter constraints than peak counts, previous CNN models, and scattering transform for all noise levels. Using MSE loss, we achieve comparable constraining power to \cite{Ribli_2019} who also used MSE loss, but used a different CNN architecture and applied smoothing to the fields. This suggests that the constraining power achieved by different deep models is fairly independent of model architectures, 
and it is the choice of the loss function that is more important for optimal performance. 

\begin{figure*}  
\centering
\begin{multicols}{3} 
\begin{subfigure}{\columnwidth}
  \includegraphics[width=1.0\linewidth]{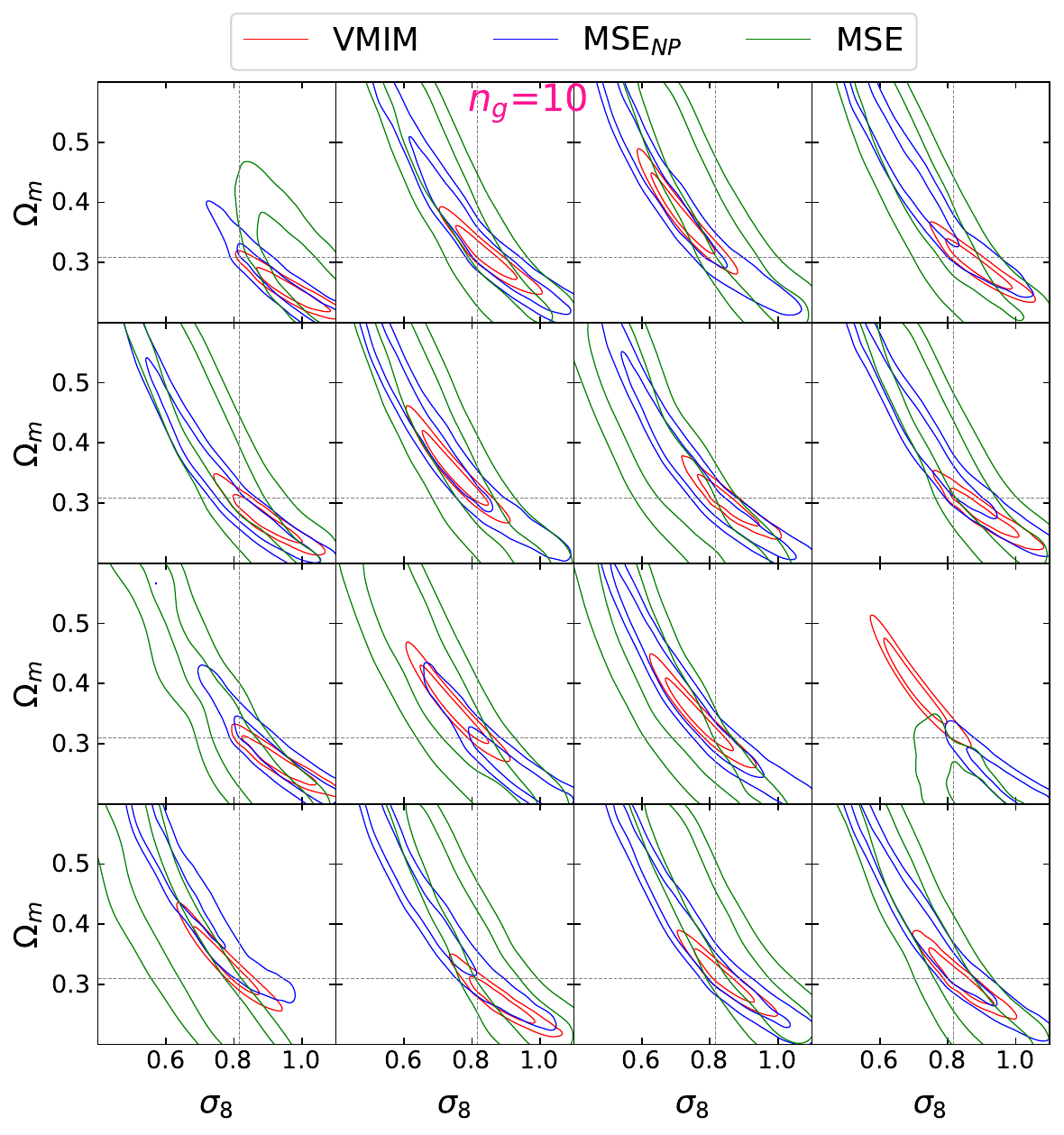}
\end{subfigure}

\columnbreak

\begin{subfigure}{\columnwidth}
  \includegraphics[width=1.0\linewidth]{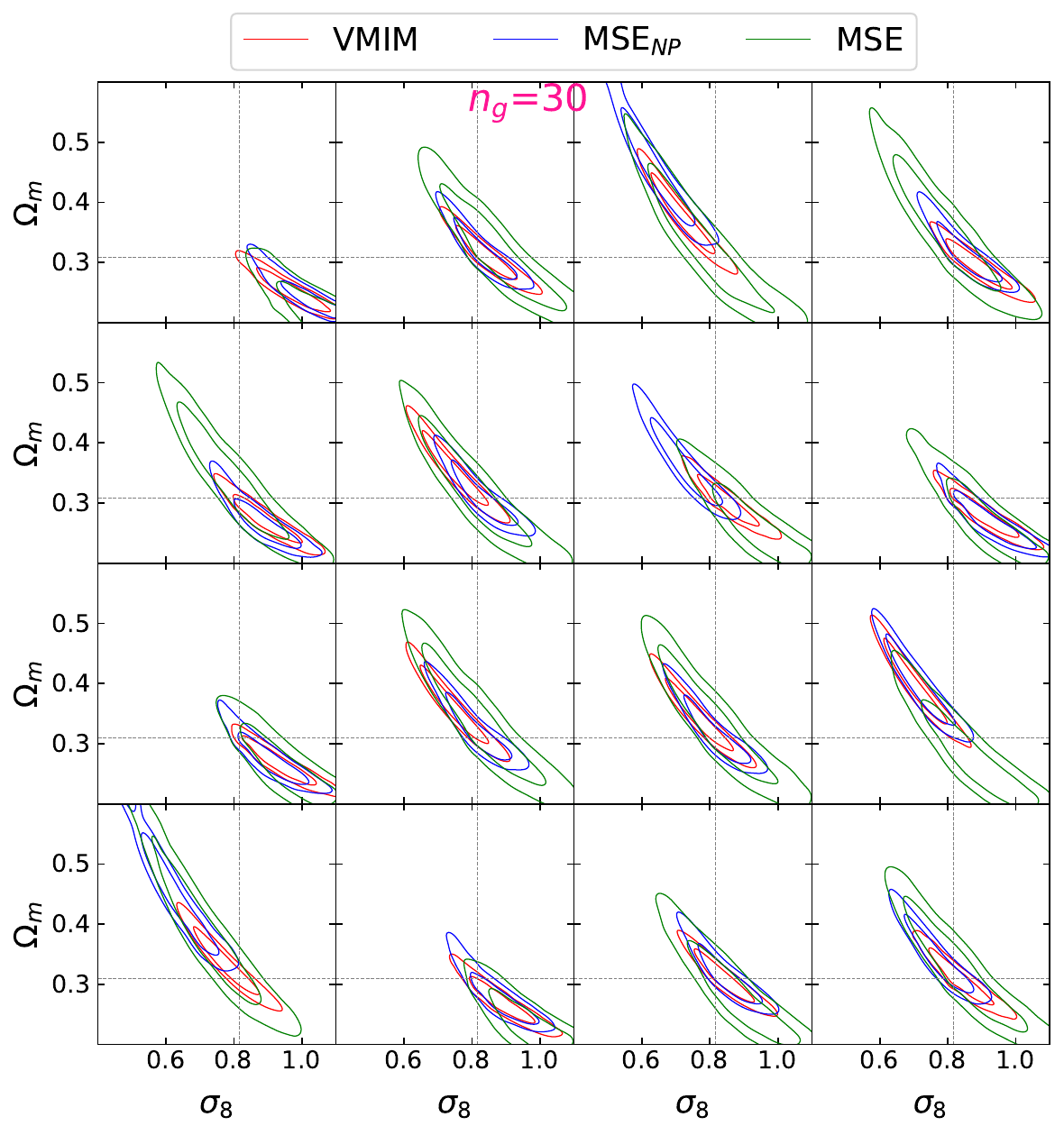}
\end{subfigure}

\columnbreak

\begin{subfigure}{\columnwidth}
  \includegraphics[width=1.0\linewidth]{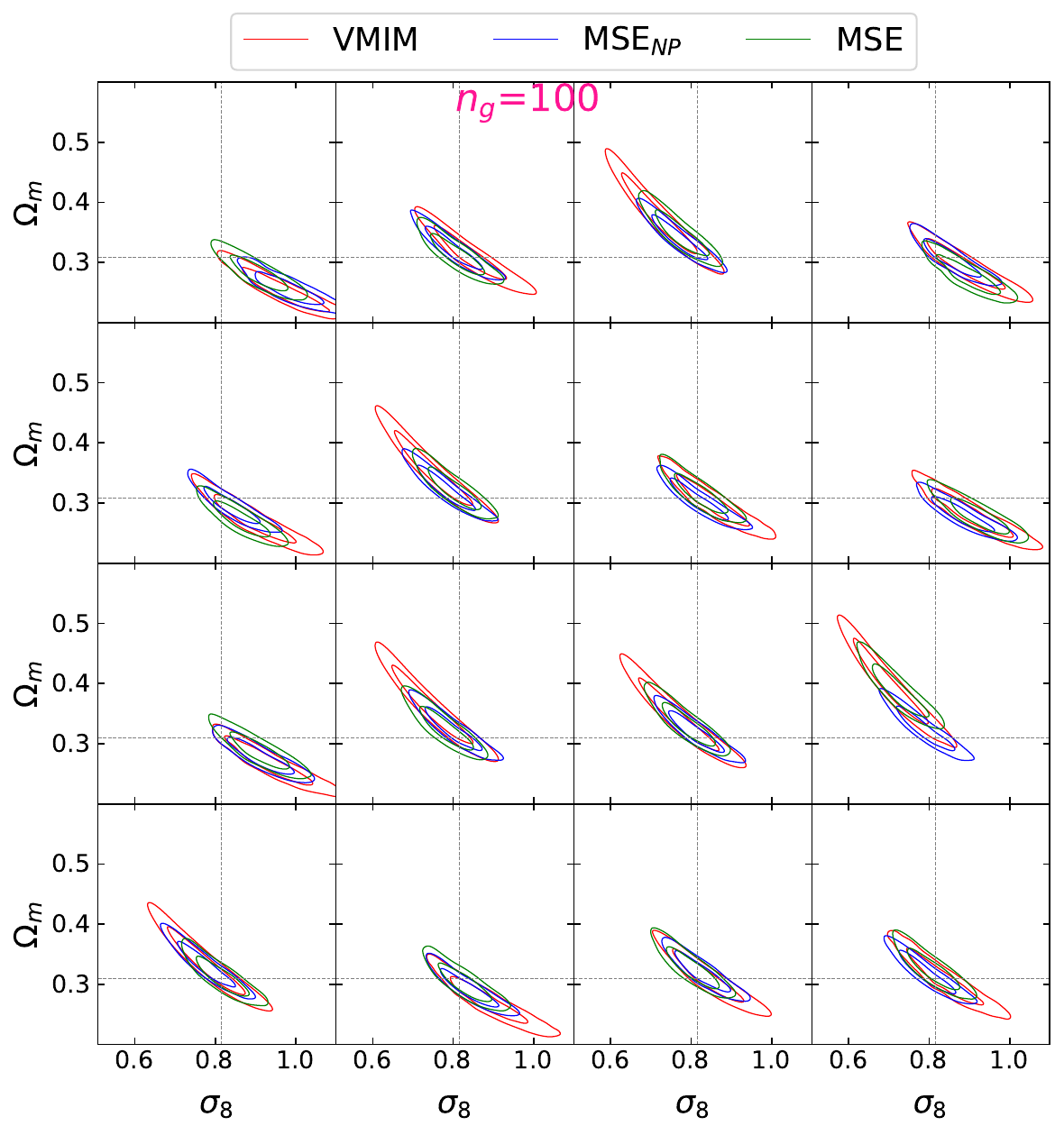}
\end{subfigure}
\end{multicols}
\caption{Comparison of the constraining power of different losses on 16 test data with the panels corresponding to galaxy number density \(n_g = 10, 30, 100 \, \text{arcmin}^{-2}\) respectively.
The posterior constraints contain the true cosmological parameters (shown in black dashed lines) for all noise levels. In correspondence with Table \ref{tab:table1}, the VMIM loss exhibits the most stringent parameter constraints, succeeded by $MSE_{\rm NP}$, $MSE_{\rm PCA}$, and $MSE$ losses, respectively. The constraining power achieved by the MSE loss is comparable to the results of \cite{Ribli_2019}. The VMIM loss achieves $\sim 2$ times smaller contours compared to the MSE loss and \cite{Ribli_2019}.}
\label{fig:RibliComparison}
\end{figure*}

To illustrate the impact of different losses on our model's performance, we present the posterior comparison of various losses using 16 test maps with galaxy number density of \(n_g = 10, 30, 100 \, \text{arcmin}^{-2}\) in Figure \ref{fig:RibliComparison}. The posterior constraints are consistent with the true cosmological parameters (shown in black dashed lines) for all noise levels. In alignment with Table \ref{tab:table1}, the VMIM loss achieves the most stringent parameter constraints, followed by $MSE_{\rm NP}, MSE_{\rm PCA}, MSE$ losses, in that order.

\begin{table*}
\centering
\caption{Comparison of the constraining power between different methods. The figure of merit is measured by the reciprocal of the 1$\sigma$ confidence area on the ($\Omega_m$, $\sigma_8$) plane, using a \(3.5 \times 3.5 \, \text{deg}^2\) convergence map. The rows with bold figures of merit correspond to results obtained using methods outlined in this study.}
\label{tab:table1}
\begin{tabular}{|l|c|c|c|}
\hline
Method & \(n_g = 10 \, \text{arcmin}^{-2}\) & \(n_g = 30 \, \text{arcmin}^{-2}\) & \(n_g = 100 \, \text{arcmin}^{-2}\) \\
\hline
ResNet18 + VMIM loss & 72 & 236 & 400 \\
ResNet34 + VMIM loss & 72 & 234 & 403 \\
ResNet50 + VMIM loss & 75 & 239 & 407 \\
ResNet18 + MSE\(_{\rm NP}\) loss & 70 & 170 & 351 \\
ResNet18 + MSE\(_{\rm PCA}\) loss  & 50 & 140 & 290 \\
ResNet18 + MSE loss  & 35 & 114 & 297 \smallskip\\
Power Spectrum & 30 & 52 & 81 \\
Peak Count & 40 & 85 & 137 \\
\cite{Ribli_2019} CNN & 44 & 121 & 292 \\
Scattering Transform \(s_0 + s_1 + s_2\) & \(\lesssim 50\) & \(\lesssim 140\) & \(\lesssim 329\) \\
Multiscale Flow ($256^2$ resolution) & 82 & 226 & 631 \\
\hline
\end{tabular}
\end{table*}

\subsection{Impact of baryons}
To study the impact of baryons on parameter constraints achieved using our models, we apply CNNs on mock weak lensing maps incorporating baryonic physics. Table \ref{tab:table2} presents the observed constraining power variations across diverse galaxy-shape noise levels. In the presence
of baryons, our models have $\sim 2.3$ times higher constraining power on cosmological parameters than the power spectrum. Using the VMIM loss, our results also outperforms previous CNN results by a factor of $\sim 1.5$.

\begin{figure*}  
\centering
\begin{multicols}{3} 
\begin{subfigure}{\columnwidth}
  \includegraphics[width=1.0\linewidth]{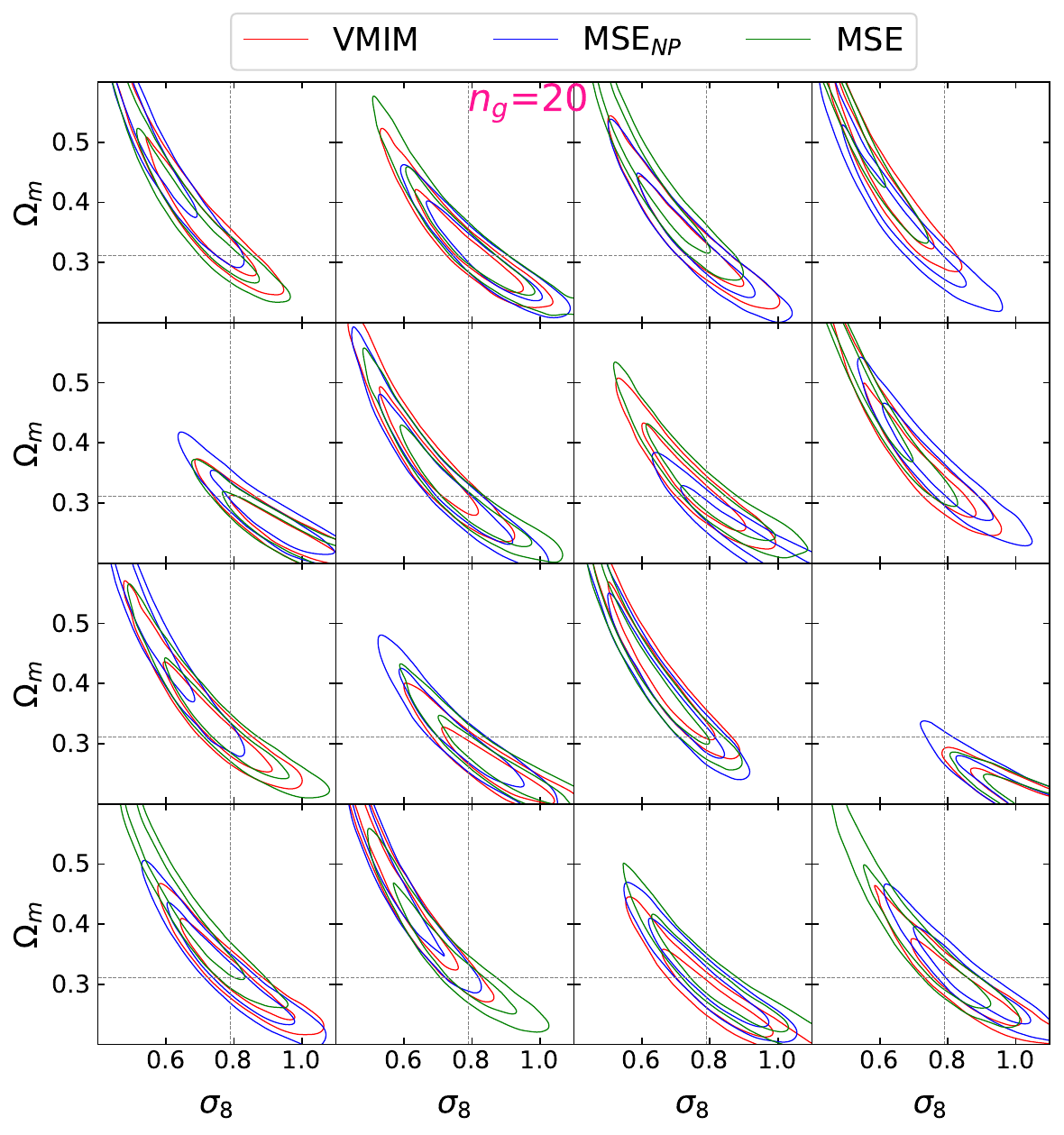}
\end{subfigure}

\columnbreak

\begin{subfigure}{\columnwidth}
  \includegraphics[width=1.0\linewidth]{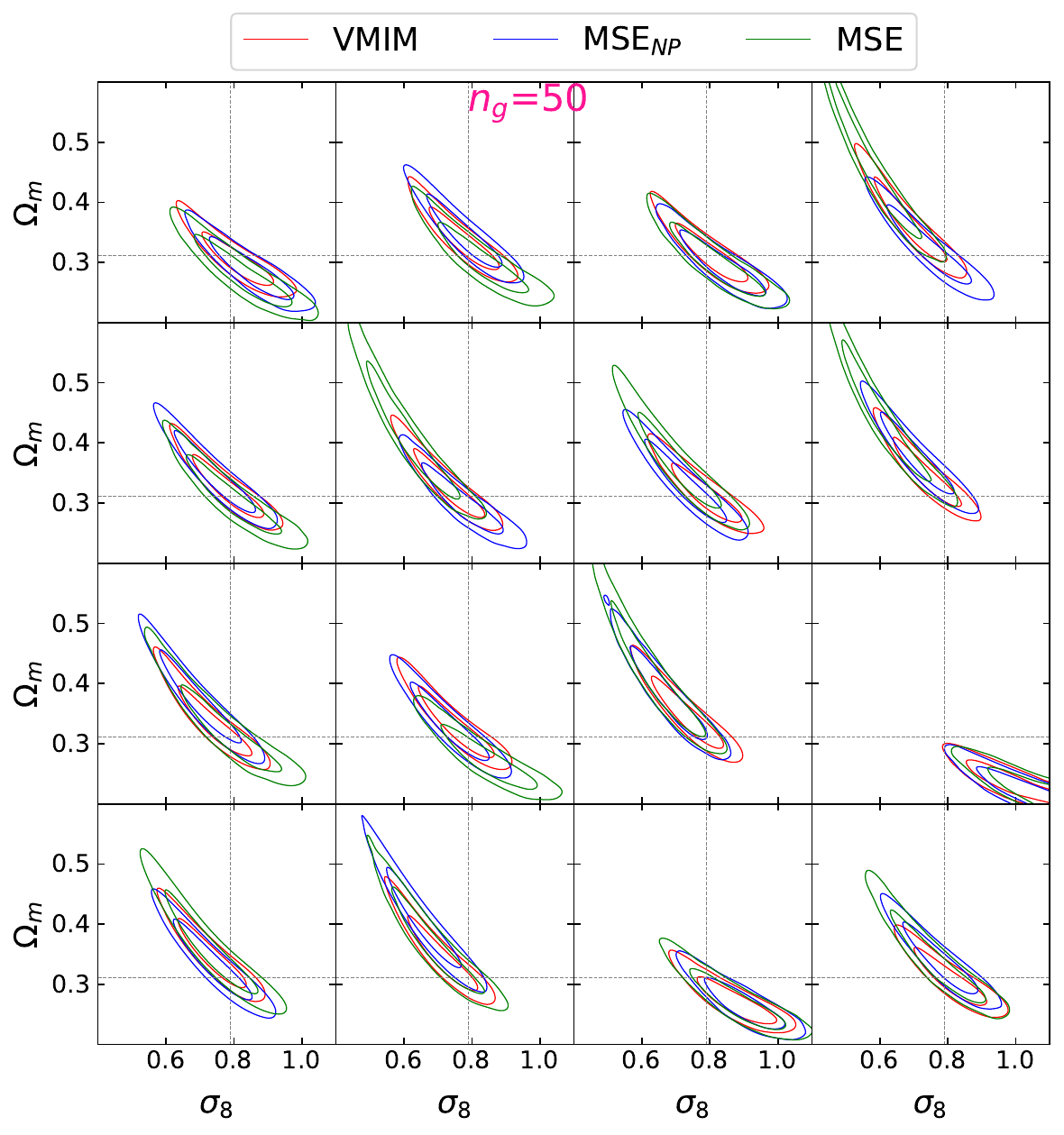}
\end{subfigure}

\columnbreak

\begin{subfigure}{\columnwidth}
  \includegraphics[width=1.0\linewidth]{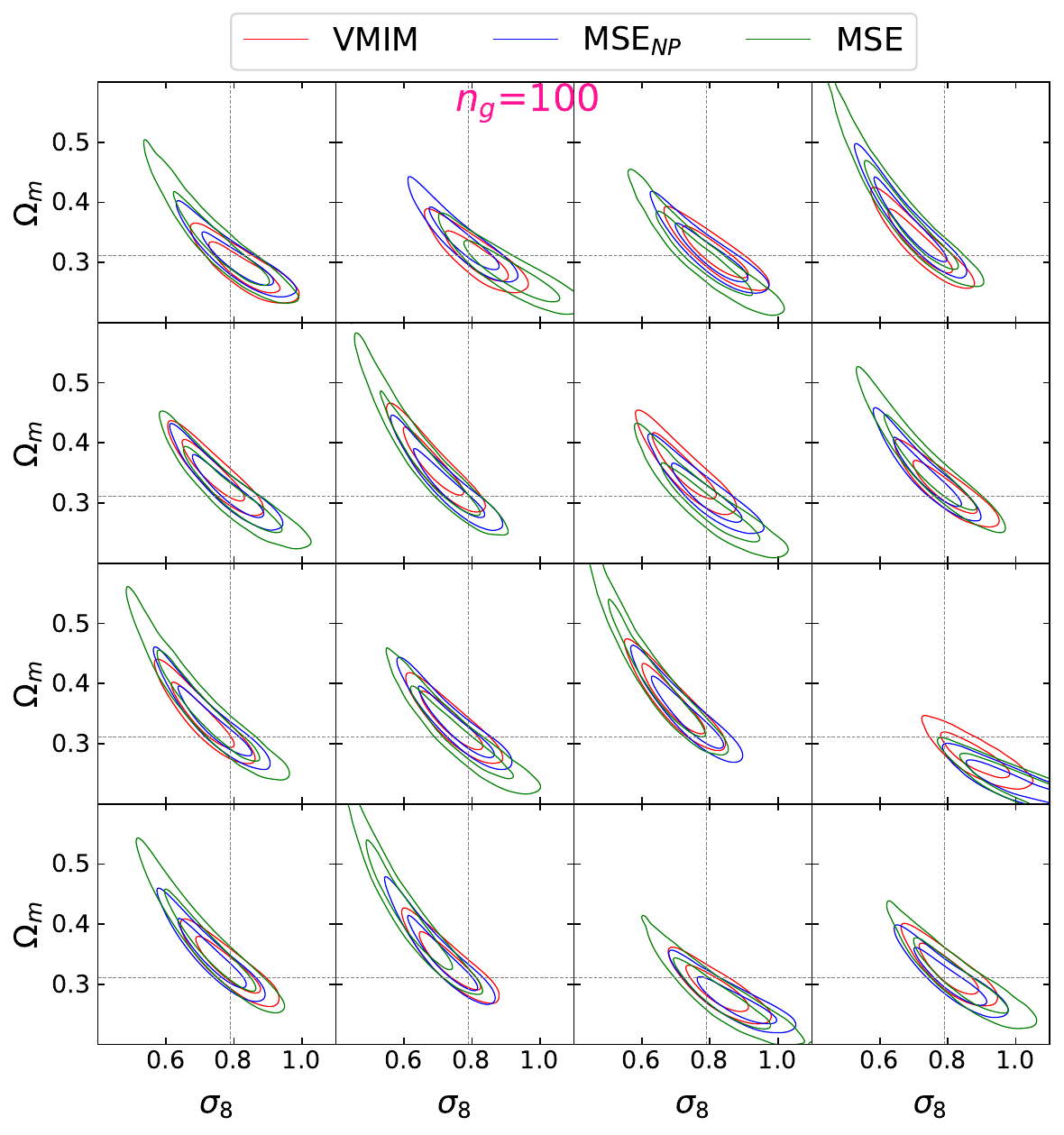}
\end{subfigure}
\end{multicols}
\caption{Comparison of the constraining power of different losses on 16 test data with the panels corresponding to galaxy number density \(n_g = 20, 50, 100 \, \text{arcmin}^{-2}\) respectively in the presence of baryons.}
\label{fig:LuComparison}
\end{figure*}

Additionally, in Figure \ref{fig:LuComparison}, we present the posterior comparison derived from assessing various loss functions using 16 test maps characterized by a galaxy number density of \(n_g = 20, 50, 100 \, \text{arcmin}^{-2}\). This comparison highlights the impact of different loss functions on the model's performance and inference outcomes.

% \begin{figure}
% \centering
% \includegraphics[width=1.0\linewidth]{plots/LuComparison.pdf} 
% \caption{Comparison of different losses on 20 test data with galaxy number density \(n_g = 20 \, \text{arcmin}^{-2}\) in the presence of baryons.}
% \label{fig:LuComparison}
% \end{figure}

\begin{table*}
\centering
\caption{Comparison of the constraining power between different methods in the presence of baryons. Here, we marginalize over baryon parameters.}
\label{tab:table2}
\begin{tabular}{|l|c|c|c|}
\hline
Method & \(n_g = 20 \, \text{arcmin}^{-2}\) & \(n_g = 50 \, \text{arcmin}^{-2}\) & \(n_g = 100 \, \text{arcmin}^{-2}\) \\
\hline
ResNet18 + VMIM loss & 112 & 151 & 202 \\
ResNet18 + MSE\(_{\rm NP}\) loss & 96 & 148 & 195 \\
ResNet18 + MSE\(_{\rm PCA}\) loss & 89 & 135 & 179 \\
ResNet18 + MSE loss  & 87 & 121 & 150 \smallskip\\
Power Spectrum & 48 & 68 & 84 \\
\cite{Lu_2022} CNN & \(\sim 77\) & \(\sim 109\) & \(\sim 136\) \\
Multiscale Flow ($256^2$ resolution) & 137 & 242 & 338\\
\hline
\end{tabular}
\end{table*}

\section{Transfer Learning}\label{sec:TL}
Transfer learning is as a Machine Learning (ML) technique where insights acquired from one task are leveraged to enhance performance in a related task \citep{Bozinovski_1976, Bozinovski_2020, Zhuang_2020}. For example, a network that has been trained to identify cars in image classification could be utilized, after fine-tuning, when attempting to identify trucks. Due to its wide prospects of application in domains where the amount of training and test data is limited, transfer learning has become an essential tool in modern machine learning research \citep[e.g.][]{Do_2005, Maitra_2015, Kabir_2022}. 

Transfer learning (TL) has also shown promise in astrophysics. \cite{Wang_2023}
demonstrated the effectiveness of transfer learning in classifying stellar light curves, enhancing the accuracy of categorizing astrophysical phenomena and understanding of stellar behavior and properties. \cite{Vilalta_2018} highlighted solutions that transfer learning introduces, particularly in dealing with the diverse range of data types encountered in astronomy.

We use three different datasets for getting our pre-trained models. First, we harness models pre-trained on the extensively benchmarked ImageNet dataset, known for its comprehensive compilation of image classification challenges. Next, we pre-train our models on gaussian random fields — a cost-effective option due to their straightforward production process. These models are then fine-tuned on our WL maps. Finally, we extend our approach to include models initially trained on log-normal fields, and these models also undergo subsequent refinement on our WL maps.

By leveraging a pre-trained neural network, our CNNs are endowed with a rich set of features and insights garnered from extensive datasets, which are then intricately honed to suit the nuances of our domain-specific task of cosmological inference. The fine-tuning phase is critical; it enables the pre-existing neural architecture to recalibrate, aligning its learned patterns and intricacies with the unique characteristics of our WL maps.
Our results of the constraining powers achieved using transfer learning are shown in tables \ref{tab:table3} and \ref{tab:table4}. %These results underscore the potential of transfer learning in enhancing the predictive prowess of discriminative methods within the context of cosmological parameter estimation.

\subsection{ImageNet dataset TL}
We used publicly available ResNet18 models pre-trained on the ImageNet dataset \citep{He_2015}. The validation accuracy on the ImageNet dataset of these models is $\sim 90\%$. The pre-trained models were then fine-tuned on our DM-only WL dataset, and parameter constraints were obtained using the same methodology as training from scratch.

Despite the high validation accuracy on ImageNet, the final models did not attain a level of performance on our WL dataset that was equivalent to that of models trained from scratch with randomly initialized weights. This is not too surprising because of the stark differences in visual features between the everyday objects of the ImageNet dataset and the complex cosmological and astrophysical patterns present in WL maps.
% ; for visual comparison, sample images from the ImageNet dataset and our WL maps are shown in figure \ref{fig:ImageNet_WL_comparison}.

% \begin{figure}
% \centering
% \includegraphics[width=1.0\linewidth]{plots/ImageNet_WL_comparison.pdf} 
% \caption{Visual comparison of sample images from the ImageNet dataset and our DM-only WL maps dataset. The images are significantly different, and so pre-training on ImageNet images yields worse parameter constraints compared to training from scratch on the WL maps.}
% \label{fig:ImageNet_WL_comparison}
% \end{figure}

\subsection{Pre-training with Gaussian random fields}
\label{subsec:GRF}
Next, we consider pre-training the networks on Gaussian Random Fields (GRFs) and test if it helps reduce the overfitting and improves the constraining power. During the pre-training, we randomly sample cosmological parameters from the prior and generate GRFs on the fly with the same power spectra as the WL maps. 

We first directly apply this pre-trained network for inference without any fine-tuning on numerical simulations. Note that this network only serves as a data compressor. As long as we measure the mean and covariance matrix in Equation \ref{eq:likelihood} with realistic WL maps, the posterior inference won't be biased even if we train the network on GRFs. In this case, we find comparable constraining power as the power spectrum analysis, which is not surprising since the GRFs only contain the power spectrum information.

After fine-tuning the networks on DM-only WL maps, we present the figure of merit in Table \ref{tab:table3}. We find comparable constraining power compared to training the models from scratch. Pre-training the model with GRFs did not improve our posterior constraints. 

%For the same reasons as using log-normal maps, we also used Gaussian random maps for pre-training our ResNet. 

%Gaussian random maps are obtained by scaling a white noise field by the square root of the required power spectrum in Fourier space \citep{Boruah_2022}. We followed the same strategy for generating the Gaussian random maps as for generating the log-normal maps outlined above.
%These maps were also produced on the fly during training, and training was done until the training loss converged, followed by fine-tuning on DM-only WL maps. 

%Using Gaussian random maps for transfer learning, we achieve comparable constraining power as training the models from scratch, as seen in table \ref{tab:table3}, suggesting that the statistics learned from the Gaussian random maps are overruled by the statistics learned from the DM-only WL maps.

\begin{table*}
\centering
\caption{Comparison of the constraining power between different methods using Gaussian random maps transfer learning. The results are similar to the results in table \ref{tab:table1}.}
\label{tab:table3}
\begin{tabular}{|l|c|c|c|}
\hline
Method & \(n_g = 10 \, \text{arcmin}^{-2}\) & \(n_g = 30 \, \text{arcmin}^{-2}\) & \(n_g = 100 \, \text{arcmin}^{-2}\) \\
\hline
ResNet18 + VMIM loss & 70 & 221 & 412 \\
ResNet18 + MSE\(_{\rm NP}\) loss & 74 & 154 & 360 \\
ResNet18 + MSE\(_{\rm PCA}\) loss  & 55 & 142 & 281 \\
ResNet18 + MSE loss  & 45 & 128 & 286 \\
%\cite{Ribli_2019} CNN & 44 & 121 & 292 \\
\hline
\end{tabular}
\end{table*}

\subsection{Pre-training with log-normal fields}
While GRFs provide a reasonable approximation of WL maps on large scales, they lack non-Gaussian features, and pre-training on these GRFs does not help with the modeling of small-scale non-Gaussian structures. We further explore pre-training with log-normal fields, which are a better approximation to the non-Gaussian WL maps and are widely used for their modeling \citep{Taruya_2002, Xavier_2016, Clerkin_2017}.

%non-Gaussian fields such as WL maps are commonly modeled as lognormal fields \citep{Taruya_2002, Clerkin_2017}. While lognormal fields do not approximate the 3D density field well \citep{Klypin_2018}, they accurately describe the 2D convergence field \citep{Clerkin_2017, Xavier_2016}.

%The convergence field of the \(i\)-th redshift bin, \(\kappa^i\), is an exponential transformation of a Gaussian field, \(y_i\) \citep{Boruah_2022}. The two variables are related via:

The log-normal fields $\kappa_{\mathrm{LN}}$ can be generated efficiently with
\begin{align}
\kappa_{\mathrm{LN}}(\theta) = e^{\kappa_{\mathrm{GRF}}(\theta)} - \lambda,
\end{align}
where $\kappa_{\mathrm{GRF}}$ is a GRF, and \(\lambda\) is referred to as the shift parameter of the lognormal distribution. The shift parameter depends on the scale at which the field is smoothed or pixelized. In this work, we measure $\lambda$ from the mock WL maps. The correlation function of $\kappa_{\mathrm{LN}}$ is related to the correlation function of $\kappa_{\mathrm{GRF}}$ via \citep{Xavier_2016}:
\begin{align}
\label{eq:corr}
\xi_{\mathrm{GRF}}(\theta) = \log\left[\frac{\xi_{LN}(\theta)}{\alpha^2} + 1\right],
\end{align}
where $\alpha=\exp{(\mu+\sigma^2/2)}$, with $\mu$ and $\sigma^2$ being the mean and variance of $\kappa_{\mathrm{GRF}}$. 

Similar to section \ref{subsec:GRF}, during pre-training, we randomly sample cosmological parameters and generate log-normal fields on the fly with the same power spectrum as the simulated WL maps. The power spectrum of $\kappa_{\mathrm{GRF}}$ is calculated numerically from Equation \ref{eq:corr}. Then we fine-tune the pre-trained network on the simulated WL maps and perform posterior inference following the same procedure described in section \ref{sec:inference}.

In table \ref{tab:table4} we show the figure of merit of posterior constraints with networks pre-trained on log-normal maps. Again, we find comparable performance as compared to training the network from scratch. This is possibly due to the specific non-Gaussian features inherent in the weak lensing data that are not fully captured by the log-normal approximation. %After all, the Fisher information of the log-normal field is the same as the GRF.

\begin{table*}
\centering
\caption{Comparison of the constraining power between different methods using lognormal maps transfer learning. The results are similar to the results in table \ref{tab:table1}.}
\label{tab:table4}
\begin{tabular}{|l|c|c|c|}
\hline
Method & \(n_g = 10 \, \text{arcmin}^{-2}\) & \(n_g = 30 \, \text{arcmin}^{-2}\) & \(n_g = 100 \, \text{arcmin}^{-2}\) \\
\hline
ResNet18 + VMIM loss & 65 & 240 & 396 \\
ResNet18 + MSE\(_{\rm NP}\) loss & 67 & 165 & 347 \\
ResNet18 + MSE\(_{\rm PCA}\) loss  & 48 & 130 & 295 \\
ResNet18 + MSE loss  & 50 & 118 & 300 \\
%\cite{Ribli_2019} CNN & 44 & 121 & 292 \\
\hline
\end{tabular}
\end{table*}

\section{Conclusions}\label{sec:Conclusions}
In this paper, we trained convolutional neural networks to constrain the underlying cosmological parameters ($\sigma_8$, $\Omega_m$) of simulated weak lensing convergence maps (with and without baryonic effects) in the presence of shape noise levels corresponding to ongoing and future large weak lensing surveys.

Our results indicate that discriminative CNN models, when trained using Mutual Information Maximization-based and principal component analysis-enhanced loss functions, yield significantly tighter constraints on the $\Omega_m - \sigma_8$ parameter space than conventional methods such as power spectrum analysis, peak counting, and CNN models trained with MSE loss. 

Notably, the Variational Mutual Information Maximization (VMIM) loss function leads to an improvement in parameter constraints, using DM-only WL maps, by factors of $\sim 5$ over the power spectrum, $\sim 3$ over peak counts, and $\sim 2$ over previous CNN models, even under varying noise levels representative of surveys like LSST and Euclid.
Using noisy WL convergence maps with baryons, we achieve $\sim 2.3$ times stronger constraining power than the power spectrum and $\sim 1.5$ over previous CNN models.

To further explore the possibilities of improvement for this task, we also discussed transfer learning where we adapted pre-trained models, also called Foundation Models, trained on a large number of different tasks or datasets, for cosmological inference. However, our results suggest that the direct training on weak lensing maps still holds a slight edge, and pretraining on a large number of Gaussian or log-normal maps provided no advantage. This is possibly due to the specific non-Gaussian features inherent in the weak lensing data that are not fully captured by the log-normal or Gaussian random approximations. This suggests that 
the use of Foundation Models may 
be of limited use in some scientific 
tasks, specially when the Foundation Model training data differ from the real 
data in some important aspects and there is enough realistic training data available so that there is no need for Transfer Learning. 

In previous work generative MultiScale Flow (MSF) \citep{dai2023multiscale}
outperformed discriminative CNN, but for realistic noise levels such as 10 or 30 $\mathrm{arcmin}^{-2}$ we find that our CNN results are comparable, potentially 
suggesting both may have extracted the 
full information content from 
the training data. For optimistic noise levels like 50 or 100 $\mathrm{arcmin}^{-2}$, MSF outperforms CNN by up to 60\% in Figure of Merit. Assuming they both 
give equal performance the choice of 
the method will thus depend on  
other requirements: discriminative CNN is simpler to 
train, and scales well to the higher 
dimensional data, while generative 
Normalizing Flow models also provide
additional tests of unknown systematics such as density estimation of the 
data (a generalized goodness of fit test). We expect that future 
applications on simulated and real data will continue to explore both of these
approaches to extract maximal 
amount of information from the data, and to assess its robustness against systematics and other effects.

\section*{Acknowledgements}
We thank the \href{http://columbialensing.org}{Columbia Lensing group} for making their simulations available and 
Francois Lanusse for useful 
discussions and software.

%%%%%%%%%%%%%%%%%%%% REFERENCES %%%%%%%%%%%%%%%%%%
\bibliographystyle{JHEP}

\bibliography{references} % if your bibtex file is called example.bib

% Don't change these lines
%\bsp	% typesetting comment
%\label{lastpage}
\end{document}